\documentclass[12pt]{article}
\renewcommand{\thefootnote}{\fnsymbol{footnote}}
\newcommand{\EQ}{\begin{equation}}
\newcommand{\EN}{\end{equation}}
\newcommand{\bea}{\begin{eqnarray}}
\newcommand{\ena}{\end{eqnarray}}
\newcommand{\vs}[1]{\vspace{#1 mm}}

\newcommand{\uda}{\nearrow \kern-1em \searrow}

\newcommand{\PL}[1]{Phys.\ Lett.\ {\bf #1}}

\newcommand{\beq}{\begin{equation}} 
\newcommand{\eeq}{\end{equation}}
\newcommand{\beqs}{\begin{eqnarray}} 
\newcommand{\eeqs}{\end{eqnarray}}
\makeatletter
\def\eqnarray{%
 \stepcounter{equation}%
 \let\@currentlabel=\theequation
 \global\@eqnswtrue
 \global\@eqcnt\z@
 \tabskip\@centering
 \let\\=\@eqncr
 $$\halign to \displaywidth\bgroup\@eqnsel\hskip\@centering
 $\displaystyle\tabskip\z@{##}$&\global\@eqcnt\@ne
 \hfil$\displaystyle{{}##{}}$\hfil
 &\global\@eqcnt\tw@$\displaystyle\tabskip\z@{##}$\hfil
 \tabskip\@centering&\llap{##}\tabskip\z@\cr}
\makeatother

\begin{document}

\begin{titlepage}
\setcounter{page}{0}
\begin{flushright}
EPHOU 02-008\\
November 2002\\
\end{flushright}

\vs{6}
\begin{center}
{\Large  Perturbative Derivation of Exact Superpotential for Meson Fields from Matrix Theories with One Flavour}

\vs{6}
{\large
Hisao Suzuki}\\
\vs{6}
{\em Department of Physics, \\
Hokkaido
University \\  Sapporo, Hokkaido 060 Japan\\
hsuzuki@phys.sci.hokudai.ac.jp} \\
\end{center}
\vs{6}

\centerline {{\bf {Abstract}}}
We derive the superpotential of gauge theories having matter fields in the
fundamental representation of gauge fields by using the method of Dijkgraaf and Vafa. We treat the theories with one flavour and reproduce a
well-known non-perturbative superpotential for meson field.
\end{titlepage}
\newpage
\renewcommand{\thefootnote}{\arabic{footnote}}
\setcounter{footnote}{0}


Recently, Dijkgraaf and Vafa~\cite{DV} proposed a general prescription for computing superpotentials via planar diagrams of matrix
theories. Several tests have been achieved for this conjecture~\cite{Chekhov}-~\cite{Gorski} mainly for
theories with potentials for the fields in the adjoint representation of the gauge group. A recent paper~\cite{DGLVZ} proved a
correspondence for matter fields in the adjoint representation. It is interesting to obtain effective superpotentials for fields
with fundamental representations since the non-perturbative superpotentials for these fields are well known (See, for example, Ref.~\cite{IS}). A
typical example is the theories for
$N_f< N_c$. In this case, a unique superpotential is known as~\cite{DDS,ILS,Int,IS}
\bea
W_{NP}=(N_c-N_f)(\frac{\Lambda^{3N_c-N_f}}{det\tilde{Q}Q})^{1/(N_c-N_f)},\label{NPpotential}
\ena
where $\tilde{Q}, Q$ fields are the fields in the fundamental representation of gauge fields.

Recently, R. Argurio, V.Campos, G.Ferretti and R.Heise\cite{ACFH} evaluated the superpotential for $N_c=2$ and $N_f=1$ and compare it to the field
theoretical results. They found the agreement up to sixth-order of the coupling constant. It is striking that we can reproduce the well-known
superpotentials from the matrix technique.

In this short paper, we will show that matrix results reproduce the non-perturbative superpotentials (\ref{NPpotential}) as well as tree level
potentioal for
$N_f=1$. We will mainly follow the notation of ref.\cite{ACFH}.

A prescription by Dijkgraaf and Vafa~\cite{DV}  is the followings.
For computing the effective 
superpotential for the glueball
field $S = -{1\over 32 \pi^2}\mathrm{tr} W^\alpha W_\alpha$
in the case of a $U(N_c)$ gauge group and 
chiral fields $\Phi_i$ in the adjoint representation interacting with a
tree level superpotential $ W_{tree}(\Phi_i, \lambda_a)$, 
we need to compute the
matrix integral to leading order in $N_c$:
\beq
    e^{-\frac{N_c^2}{S^2} {\cal F}_{\chi=2}(S, \lambda_a)} \approx
    \int\;d\Phi_i\; e^{-\frac{N_c}{S}  W_{tree}(\Phi_i, \lambda_a)},
\eeq
where we have denoted by $\lambda_a$ the coupling constants appearing in 
the superpotential.
The effective superpotential is, in this case \cite{DV}:
\beq
    W_{DV}(S, \Lambda, \lambda_a) = N_c S (-\log(S/\Lambda^3) + 1) +
       N_c \frac{\partial {\cal F}_{\chi=2}(S, \lambda_a)}{\partial S},
    \label{VYgen}
\eeq
where the presence on $N_c \frac{\partial}{\partial S}$ is justified by the
combinatorics of diagrams written on surfaces with spherical topology.
The first piece of the superpotential is the Veneziano-Yankielowicz
superpotential 
for pure $SU(N_c)$ SYM \cite{VY}, while the second piece
which starts with $O(S^2)$ terms gives the instantonic corrections.
In the case that the matrix model is integrable we can write the
exact effective superpotential in closed form, otherwise, we can 
compute it at any given order in $S$. Recent checks and developments
of the conjecture have been performed in \cite{Chekhov}--\cite{Gorski}.

It is natural to extend the conjecture to theories including
one matter chiral multiplets in the fundamental representation of $U(N_c)$.
This is implemented simply by including surfaces with boundaries.
To be specific, in the case of gauge group $U(N_c)$, for adjoint matter
$\Phi_i$ and fundamental matter $Q$ and $\tilde Q$ one should first
compute
\beq
    e^{-\frac{N_c^2}{S^2} {\cal F}_{\chi=2}(S, \lambda_a) -
       \frac{N_c}{S} {\cal F}_{\chi=1}(S, \lambda_a)} \approx
    \int\;d\Phi_i\; d Q\; d \tilde Q\; 
       e^{-\frac{N_c}{S}  W_{tree}(\Phi_i, Q, \tilde Q,\lambda_a)},
\eeq
and then write (the non-orientable contribution $G_{\chi=1}(S, \lambda_a)$
is absent in this case):
\beq
    W_{DV}(S, \Lambda, \lambda_a) = N_c S (-\log(S/\Lambda^3) +1) + 
       N_c \frac{\partial {\cal F}_{\chi=2}(S, \lambda_a)}{\partial S} +
        {\cal F}_{\chi=1}(S, \lambda_a).
\eeq

We can easily evaluate the superpotential. We take
a $U(N_c)$ gauge theory with one adjoint chiral multiplet $\Phi$ and one
chiral multiplets in the fundamental $Q$ and $\tilde Q$.
The tree level superpotential gives masses to all matter fields, and
moreover there is a cubic coupling between the fundamentals
and the adjoint. Other possible couplings are turned off.
The tree level superpotential is
\beq
W_{tree}={1\over 2}M \mathrm{tr} \Phi^2 + m Q \tilde Q  
+ g Q\Phi \tilde Q  ,
\eeq
where the color indices are not
written explicitly.

Since there are no self interactions of the adjoint field $\Phi$, all
diagrams will involve at least one flavor loop that is a boundary.
As a consequence, the genus zero piece of the matrix integral vanishes,
${\cal F}_{\chi=2}=0$. To leading order in $N_c$, the matrix integral is 
thus saturated
by planar diagrams with one boundary, which sum up to ${\cal F}_{\chi=1}$.

The matrix integral can thus be easily performed. We can write
\beq
Z=\int\;d\Phi_i\; d Q\; d \tilde Q \; e^{-{N_c\over S}\left(
{1\over 2}M \mathrm{tr} \Phi^2 + m Q \tilde Q  
+ g Q\Phi \tilde Q \right)}
= \langle e^{-{N_c\over S}g Q\Phi \tilde Q } \rangle,
\eeq 
where the correlators are normalized in the form:
\beq
\langle Q_{\alpha }\tilde Q^{\beta }\rangle = {1\over m}{S\over N_c}
\delta^\beta_\alpha , \qquad \qquad 
\langle \Phi^\alpha_\beta \Phi^\gamma_\lambda \rangle = {1\over M}{S\over N_c}
\delta^\alpha_\lambda \delta^\gamma_\beta.
\eeq
Expanding the exponential, we get
\beq
\langle e^{-{N_c\over S}g Q\Phi \tilde Q } \rangle = 
\sum_{k=0}^\infty {1\over (2 k)!} \left({g N_c \over S}\right)^{2k}
\langle (Q\Phi \tilde Q )_1 (Q\Phi \tilde Q )_2 \dots (Q\Phi \tilde 
Q )_{2k}\rangle, \label{expan}
\eeq
where we took into account that only correlators of an even number 
of fields $\Phi$ are non zero.

It is now a simple combinatorial problem to extract from (\ref{expan})
the coefficients of the connected planar diagrams with one boundary.
The different diagrams can be obtained first by contracting
the $Q$s and $\tilde Q$s in $(2k-1)!$ ways to give a single boundary,
and then connecting $2k$ points on the boundary through $k$ non intersecting
lines (the $\langle \Phi\Phi\rangle$ propagators).
The solution to this last combinatorial problem can be found in \cite{BIPZ},
Eq.~(31). The result for the free energy is:
\beq
\frac{N_c}{S} {\cal F}_{\chi=1}(S, g,m,M)=- \sum_{k=1}^\infty
{(2k-1)!\over (k+1)! k!} \left({g N_c \over S}\right)^{2k}
\left({S\over m N_c}\right)^{2k} \left({S\over M N_c}\right)^k N_c^{k+1},
\eeq
which we can rewrite, for $\alpha={g^2\over m^2 M}$, as:
\beq
{\cal F}_{\chi=1}(S, \alpha)=- \sum_{k=1}^\infty 
{(2k-1)!\over (k+1)! k!}
\alpha^k S^{k+1}.
\eeq
This expression can actually be summed to give:
\beq
{\cal F}_{\chi=1}(S,\alpha)=- S \left[
{1\over 2} +{1\over 4\alpha S}(\sqrt{1-4\alpha S}-1) - \log\left(
{1\over 2}+{1\over 2}\sqrt{1-4\alpha S}\right) \right].
\eeq
In this way, we can obtain the superpotential of the form:
\beq
W_{DV}=N_c S (-\log(S/\Lambda^3) +1)+{\cal F}_{\chi=1}(S,\alpha)
\label{exact}
\eeq
which is expected to be the exact superpotential for our theory with one flavor
and the Yukawa coupling to the adjoint matter field. This is a result obtained in Ref.~\cite{ACFH}.
The above form of superpotential seems rather complicated. Therefore, some simplification must be required.

We are now going to rewrite to the superpotential for meson fields.
The equation of the gluino fields can be obtained from $\partial_S W_{DV}=0$:
\beq
W'/\Lambda^3=-\log{[(S/\Lambda^3)^{N_c}/(1+\sqrt{1-4\alpha S})/2]}=0
\eeq
which implies
\newcommand{\tildes}{(S/\Lambda^3)}
\beq
\tildes^{2N_c}-\tildes^{N_c}+a\tildes=0,\label{EofM}
\eeq
where $a=\alpha \Lambda^3$.
Using this relation, We can rewrite the superpotential in the form
\beq
W_{DF}=(N_c-1)S+\Lambda^{3N_c}S^{-N_c+1}-\frac{g^2}{2m^2M}\Lambda^{6N_c}S^{-2N_c+2}.
\eeq
In this form of the potential, we can obtain the equation of motion (\ref{EofM}) and the same value as the original form of the potential when all
fields are integrated out. In order to recover a matter field, we use a matching relation$\Lambda^{3N_c}=m\tilde \Lambda^{3N_c-1}$ and write the
superpotential in the form:
\beq
W_{DF}=(N_c-1)S+m\tilde \Lambda^{3N_c}S^{-N_c+1}-\frac{g^2}{2M}\tilde \Lambda^{6N_c-2}S^{-2N_c+2}.
\eeq
The expectation value of $X=Q \tilde Q$ fields can be given by $X=\partial_m W_{DV}$, which implies
\beq
X=\tilde \Lambda^{3N_c-1}S^{-N_c+1}
\eeq
By using this fields, we can write the superpotential as
\beq
W_{DV}=(N_c-1)(\frac{\tilde \Lambda^{3N_c-1}}{X})^{1/(N_c-1)}+mX-\frac{g^2}{2M}X^2.\label{final}
\eeq

Now we can easily interpret the origin of each term of the superpotential. Consider the $U(N_c)$ theory with
an adjoint matter field $\Phi$ and $N_f=1$ chiral fields
in the fundamental $Q$ and $\tilde Q$. 

The tree level superpotential is given by
\beq
W_{tree}={1\over 2}M \mathrm{tr} \Phi^2 + m Q \tilde Q 
+ g Q\Phi \tilde Q. \label{ftree}
\eeq
Let us first integrate out the adjoint field $\Phi$.
This trivially gives $\Phi=-{g\over M}\tilde Q Q$, and substituting
into (\ref{ftree}) we find,
\beq
W_{tree}=mX - {1\over 2}{g^2\over M} X^2.
\eeq

The first term in (\ref{final}) represents the exact superpotential obtained in\cite{DDS,ILS}:
\bea
W_{NP}=(N_c-N_f)(\frac{\Lambda^{3N_c-N_f}}{det\tilde{Q}Q})^{1/(N_c-N_f)},\label{famous}
\ena
which can be written in the case $N_f=1$ as
\bea
W_{NP}=(N_c-1)(\frac{\hat \Lambda^{3N_c-1}}{X})^{1/N_c-1}
\ena
Therefore, DV superpotential reproduces the famous superpotential (\ref{famous}) for $N_c=1$ as well as the tree level potential(\ref{ftree}).
Namely,
\bea
W_{DV}=W_{NP}+W_{tree}
\ena

Note that our transformation can be justified at the level of the superpotential integrated out all the fields. Namely, we have identified
the superpotential  when the superpotentials are written in the variables $\Lambda, m, g$ and $M$. However, we can easily
integrate in the original fields by the usual procedure.

In this paper, we have shown that the superpotentials for meson fields can be derived from matrix technique. We must investigate whether "Large
N" or just a "planar" for the matrrix-superpotential correspondence.

Lastly, we are going to evaluate superpotentials explicitly.
First of all, we choose a variable $y=(\frac{m}{\Lambda^3}X)^{1/(N_c-1)}$. Then the superpotential can be written as
\bea
W/\Lambda^3=(N_c-1)y^{-1}+y^{N_c-1}-\frac{1}{2}ay^{2N_c-2},
\ena
The equation of motion can be written in the form
\bea
f(y)=1-y^{N_c-1}+ay^{2N_c-1}=0
\ena
As a vacuum solution, we choose a solution around $y=1$ for small value of $a$. Note also that $W'(y)=-(N_c-1)\Lambda^3f(y)/y^2$
The value of superpotential $W$ at the pole of $f(y)$ can be evaluated as
\bea
W=\oint\frac{dy}{2\pi i}W(y)\frac{f'(y)}{f(y)},
\ena
where the contour encircles a zero of the function $f(y)$.
By performing partial integration, we can write
\bea
W=(N_c-1)\Lambda^3\oint\frac{dy}{2\pi i}y^{-2}f(y)\log f(y)
\ena
We use the integral representation for the logarithm and write
\bea
W=(N_c-1)\Lambda^3\oint\frac{ds}{2\pi i s^2} \oint\frac{dy}{2\pi i}y^{-2}f(y)^{s+1},
\ena
where the contour of $s$ encircles the point $s=0$.
Since the contour integral of the variable $y$ has a cut by the introduction of the variable $s$, We can represent it as a line integral between $0$
to a pole which shrinks to y=1 upon expanding with respect to $a$, therefore, we can write
\bea
W&=&(N_c-1)\Lambda^3\oint\frac{ds}{2\pi i s^2} \frac{\sin{\pi (s+1)}}{\pi}\int_0^1dyy^{-2}(1-y^{N_c}+ay^{2N_c-1})^{s+1}.
\ena
By changing variables to $y=t^{1/N_c}$ and expanding the superpotential with respect to $a$, we can evaluate the $t$-integration by the usual beta
integral. We also find the integral with respect to $s$ has just a single pole, and we finally get an expression of superpotential:
\bea
W &=& N_c\Lambda^3\sum_{n=0}^\infty\frac{\Gamma((2-1/N_c)n-1/N_c)/\Gamma(-1/N_c)}{\Gamma((1-1/N_c)n+2-1/N_c)/\Gamma(2-1/N_c)n!}a^n \\
{}&=& N_c\Lambda^3[1-\frac{1}{2N_c}a-\frac{N_c-1}{2N_c^2}a^2-\frac{(5N_c-4)(N_c-1)}{6N_c^3}a^3+\cdot\cdot\cdot],
\ena
The first few terms perfectly agrees with the result for $N_c=2$\cite{ACFH}.

The author would like to thank K.Suehiro for valuable discussions.


\end{document}